# X-probability and Irreversibility Paradox

V.A. Skrebnev

Physics Department, Kazan State University, Russian Federation
e-mail: vskrebnev@mail.ru

**Abstract**

Based on the general thought that mechanics cannot be absolutely exact, supported by pulsed nuclear magnetic resonance (NMR) experiments on the reversal of time in spin systems, we postulate existence of a probability which do not follow from the Schrödinger equation, or X-probability. It is shown that the X-probability allows to explain irreversibility of evolution of macrosystems, and also creates a basis for using probabilistic methods in statistical mechanics and for deliverance from the mechanical determinism.

## Introduction

Equations of mechanics, both classical and quantum, lay in the foundation of modern natural sciences. One of the essential features of these equations is reversibility in time. Hence, from the point of view of mechanics, all processes in nature should be reversible, which would mean that heat can transfer from a cold body to a hot one, as well as it transfers from hot to cold, and tomorrow would be same as yesterday. However, as we know from the experience of mankind, it does not occur, and processes in the real world appear to be irreversible.

The irreversibility problem can be placed with a good reason among the Great Problems in science. This problem has been investigated by Thomson, Maxwell, Helmholtz, Gibbs, Born, Einstein, Hopf, Neumann, Pauli, Poincare and many other famous scientists. The irreversibility problem was crucial in the work of such luminaries as Boltzmann, Planck and Prigogine.

The basic problem in the explanation of irreversibility paradox is that the reversible equations of mechanics are supposed to apply without any changes also to evolution of macrosystems, *i.e.* systems containing astronomical number of particles. However, the equations of classical mechanics appeared as a result of observations of a small number of bodies visible to a naked eye and moving with velocities much less than velocity of light. It is a little wonder that they turned out to be inaccurate in the range of velocities comparable with velocity of light, as well as for description of microcosm phenomena. Equations of quantum mechanics result from observations of a small number of microparticles, too. Couldn't they turn out to be inaccurate for describing systems with astronomically large number of particles? In other words we cannot assert categorically that the reversible equations of mechanics can be extrapolated directly on macro systems with number of particles of the order of $10^{23}$. Nonetheless, for over a hundred years up to nowadays, attempts have been made to derive irreversibility from reversible mechanical equations.

We do not aim to analyze here all «solutions» of the irreversibility problem. A careful reader is able to find himself/herself a point in every such a «well-founded solution», at which the irreversibility is introduced «by hands», quite often unconsciously. In the case of classical mechanics it applies to an "explanation" of irreversibility through instability of the system and appearance of so-called dynamical chaos in it. Behavior of the unstable system remain deterministic and reversible, while the term "chaos", applied in regard to such systems, is misbegotten and misleading. The use of a rough, coarse-grained, distribution function is not a solution either, for the dynamic equations offering no possible mechanisms for making things rough. Yet, the first thing one should bear in mind is that for description of a real system, consisting of microparticles, one should use equations of quantum mechanics. These equations have no exponentially diverging solutions, and such systems are therefore locally stable, while local instability is needed for appearance of the dynamical chaos.

In any case and under all circumstances reversible equations cannot describe irreversible processes.

Another problem is that of determinism. According to mechanics, all states of any physical system are defined by initial conditions (the state of the system at the moment of time which we accept as initial). This circumstance has occasioned the so-called mechanical or Laplace determinism according to which everything that will be is already predetermined – implying, among other things, that the choice of a wife or a president, was determined not by the freedom of our will but by some initial conditions. As it will be shown soon, having found the source of irreversibility, we not only solve the irreversibility problem, but also can be liberated from the determinism either.

## X-probability Postulate

The very fact of irreversibility of macroprocesses shows that it is incorrect to extrapolate the mechanical theory to macrosystems. It means that something is missing in the mechanics equations for description of macroprocesses. However, on the macro-level it is impossible to compare predictions of mechanics and experimental results because it is impossible to find a solution of the equations of motion for the system of $10^{23}$ particles. It would be another matter if we could reverse the sign of time in equations of mechanics. In this case return of the system into the initial state would mean the absence of irreversibility and could put the second law of thermodynamics in a difficult position. On the other hand, the failure of the system to return into the initial state would experimentally manifest the insufficiency of quantum mechanics to describe macroprocesses and put an end to the irreversibility problem.

In the foreseeable future it will hardly be possible to reverse time in the whole Universe. But for a single isolated system such procedure is possible. It is known that the common solution of the basic equation of quantum mechanics – the Schrödinger equation – looks like

$$\psi(t) = \psi(0)e^{-i\frac{\hat{\mathrm{H}}}{\hbar}t} \quad (1)$$

where $\hat{\mathrm{H}}$ is an energy operator or Hamiltonian of a system independent of time. From equation (1) it is clear that change of the Hamiltonian sign is identical to change of the time sign. We managed to perform experiments in which we changed, with predetermined accuracy, the sign of the energy operator in a system of magnetic moments placed in an external magnetic field [1, 2]. The results of these experiments cannot be correctly described on the basis of reversible equations of quantum mechanics. In fact, was it possible, it would have meant a violation of the second law of thermodynamics.

Our experiments and common sense speak for incompleteness of the quantum mechanics. In [1,2] we have proposed the existence of certain changes of system states at conserved energy, which do not follow from the Schrödinger equation.

The Schrödinger equation gives the probability of instant values of physical variables, for example, particle co-ordinates and momenta. In the quantum mechanics the concept of a particle trajectory is meaningless. Instead of moving along a certain trajectory, the particle with certain probability appears in this or that point of space (we are leaving aside the issue of what a particle is and what are the mechanisms of its appearing in space). It is reflected in the concept of identity of particles in microcosm. For example, we can say, that the atom of helium has two electrons, but it is meaningless to speak about which of them appears in this or that point of space. Identity of electrons leads to occurrence of so-called exchange energy and an origin of the two types of helium – ortohelium and parahelium.

If quantum mechanics is absolutely accurate, the probability of the instant values of physical variables would be exhaustively defined by solution of the Schrödinger equation. As no physical theory can be absolutely exact, it is natural to suppose that quantum mechanical probability does not exhaust completely the probabilistic nature of the world. Hence, it is necessary to complement the quantum mechanical probability of instant values of the physical variables with a probability which is not contained in the basic equation of quantum mechanics. We will coin it X-probability. This probability manifests itself in the random change of instant values of physical variables (random with respect to that determined by the Schrödinger equation), for example, particle co-ordinates and momenta. This randomness correlates with additional changes in the states of microparticles in the system, and hence, of the state of a system as a whole. We will coin such changes by X-jumps.

In systems with small number of particles there is no basis to question the correctness of quantum mechanical description. This implies that X-jump probability in such systems is small, but it grows with the growth of number of particles. In macrosystems these conditions are already fulfilled, that is why it is impossible to neglect X-jumps.

Let us make some estimations. Let the X-jump probability for one particle be equal to $10^{-10}$. It means that such jump occurs approximately once in 300 years.

The state function of the system of N particles can be represented as a linear combination of products of N one-particle functions. If the number of particles in the system is equal to $10^{23}$, the number of changes in system state will be $10^{13}$ per second. Therefore, in one millisecond the system will have time to go through $10^{10}$ different states, randomly distributed over the system states space. It is enough for the measured value of the physical quantity to be equal to the value averaged over the whole ensemble of the system states.

A consequence of the X-jumps is that the system cannot return to its initial state when the time sign is reversed. The evolution of the system becomes irreversible: random jumps during the system evolution cannot be repeated in the "backwards" motion simply because they were random. In fact, upon tracing back the other random jumps will happen, so the system will not return to its initial state.

**X-probability and statistical mechanics**

The irreversibility problem is closely connected with the problem of substantiation of statistical mechanics, based on the use of the probability theory (it is necessary to note that the probability in statistical mechanics has nothing to do with quantum-mechanical probability). The introduction of probabilistic assumptions to statistical mechanics is justified by the impossibility of solving the equations of mechanics for a macrosystem. These equations are supposed to be applied to macroprocesses without any changes (see any textbook on statistical mechanics).

Probabilistic assumptions, entered in statistical mechanics, deny determinism and contradict mechanics; at the same time the so-called stochastic movement equations (such as the Langevin equation or the Fokker-Planck equation), developed on the basis of the probabilistic approach, are the major equations of statistical mechanics. These equations describe an irreversible evolution which ends by a state of thermal equilibrium.

When deriving stochastic equations of movement, at each point of time, physical variables (for example, coordinate or velocity of a particle) are considered to have arbitrary value in the allowable range, which means they are random variables. It is impossible to explain randomness within the framework of mechanics, whereas X-jumps provide randomness of the physical variable values necessary for derivation of stochastic equations.

The X-jumps give rise to what is called chaotic behavior of the system. In other words, the existence of X-jumps provides the answer to the main question of statistical mechanics: where does determinism go, and why is it possible to use probabilistic methods for description of macrosystems?

The stochasticity generated by X-jumps does not contradict mechanics; it is an addition to the evolution of the system described by equations of mechanics. In particular, it substantiates Boltzmann's assumption made when he worked on his famous kinetic equation for rarefied gases, called Stosszahlansatz, or a hypothesis of molecular chaos. Unfortunately, the assumptions made by Boltzmann in his theory of gases were not accepted by the scientific community of his time. Non-recognition of the major scientific results has led the great scientist to heavy depression and, eventually, to suicide.

Boltzmann could not find a substantiation for his genius guesses within the limits of the mechanical theory. We will never know, but he might have doubted the infallibility of mechanics. But, for obvious reasons, it would have been unproductive to share such doubts with colleagues and opponents.

In statistical mechanics, observed values of physical quantities are achieved by averaging on ensemble of all system states corresponding to the given value of system energy. However the solution of the mechanics equations results in a unique system state corresponding to certain initial conditions that excludes the possibility of averaging on ensemble. This problem is also solved by the existence of X-jumps: the X-jumps allow the system to visit in a short time a great number of various states uniformly distributed over the space of states with given energy. As a result, we are justified in using statistical ensembles to find correct time average values of observables.

It is known that uniform distribution of systems of ensemble on the surface of constant energy corresponds to experience, which is reflected in the principle of equality of aprioristic probabilities. Direct application of this principle leads to construction of microcanonical ensemble. But it is more convenient mathematically to use the canonical distribution, giving probability of a system state with energy $E_{\mathrm{m}}$:

$$P_m = Z^{-1}\, e^{-\beta E_m}, \;\; Z = \sum_m e^{-\beta E_m} \;, \tag{2}$$

where $\beta = 1/k_B T$ , and $T$ is the absolute temperature of the system.

Formula (2) is the major formula of statistical mechanics and one of the most important formulas of physics as a whole. The observed value of physical quantity is achieved by the averaging over all values of energy.

When canonical distribution is derived, a system S is considered as a part of a very big system U described by microcanonical distribution (see, for example, [3]). Interaction of system S with environment W (or addition to system U) is considered, on one hand, necessary for an exchange of energy, but, on the other hand, negligibly small – to make it possible to talk about certain quantum state $m$ of the system S with the energy $E_m$. In other words, S and W are considered to be practically unconnected and not influencing each other [3].

When canonical distribution is derived traditionally, the energy of environment $E_u - E_m$ corresponds to the value of the energy of the system S equal to $E_m$. It means that $E_m$ is considered as the total energy of the system S equal to some quantum value, and the system S can appear in states with various values of $E_m$ exclusively thanks to an energy exchange with the environment.

To receive observed values of physical quantities by means of distribution (2), it is necessary for the system S to appear repeatedly in all states $m$ during observation with the probability defined by (2). It means that there should be extremely fast energy exchange with the environment, i.e. the interaction energy should be very high. However, in this case it makes no sense to say that the system S is reliably in this or that quantum state. Besides, the system, which is practically unconnected with the environment, will not take care about the state of the environment, and the assumption that the big system U is in equilibrium is not physically well-founded. On the whole we can say that the assumptions made for the traditional derivation of canonical distribution contain insurmountable internal contradictions.

Taking into account the stochasticity in behavior of a system, it is possible to derive canonical distribution without considering a system as a part of a big system in equilibrium. We will consider the system consisting of N identical particles. We will neglect the energy of interaction in the system in comparison with the energy of the system's particles. Let $n_i$ be a number of particles at the given energy level. Then, the total energy of the system is:

$$E = \sum_i n_i \, \varepsilon_i \tag{3}$$

Maintaining N particles we have:

$$\sum_i n_i = N \, . \tag{4}$$

A great variety of values of $n_i$ corresponds to condition (3). Each set $n_i$ creates a certain configuration. Taking into account permutations of particles, the given configuration can be realized in P ways:

$$P = \frac{N!}{n_1! \, n_2! \ldots n_k! \ldots} . \tag{5}$$

The maximum value of P corresponds to the system equilibrium state. Using Lagrange method of uncertain multipliers, we find that the maximum of P under additional conditions (3) and (4) is reached when $n_i$ equals to

$$n_i = e^{-\alpha} \, e^{-\beta \, \varepsilon_i} \tag{6}$$

The value $e^{-\alpha}$ is found from (4) as follows:

$$e^{-\alpha} = \frac{N}{\sum_i e^{-\beta \, \varepsilon_i}}$$

The probability of a particle to occupy the given energy level equals to

$$W_i = \frac{n_i}{N} = \frac{e^{-\beta \, \varepsilon_i}}{\sum_j e^{-\beta \, \varepsilon_j}} = \frac{e^{-\beta \, \varepsilon_i}}{Z_1} . \tag{7}$$

The product of probabilities (7) for all particles of the system is

$$W^{(k)} = \prod_1^N \frac{e^{-\beta\varepsilon_i^{(k)}}}{Z_1} = \frac{e^{-\beta E_{(k)}}}{Z_1^N}, \qquad (8)$$

where $\varepsilon_i^{(k)}$ is the energy of a particle $i$ at certain distribution of particles over levels, and $E_{(k)}$ is the energy of a system at the given distribution, gives a probability of the system state with the energy $E_{(k)}$ without taking into account permutations of identical particles.

As the product of exponents is equal to the exponent from the sum, we have

$$Z_1^N = \sum_m e^{-\beta E_{(m)}}, \qquad (9)$$

where summation occurs over all particle distributions on the levels.

Considering permutations of identical particles, we find for probability of a system state with the energy $E_{(m)}$:

$$W_{(m)} = N! \frac{e^{-\beta E_{(m)}}}{\sum_k e^{-\beta E_{(k)}}}. \qquad (10)$$

Expression (10) can be rewritten as

$$W_{(m)} = \frac{e^{-\beta E_{(m)}}}{Z}, \qquad (11)$$

where

$$Z = \frac{1}{N!} Z_1^N \qquad (12)$$

Distribution (11) is the canonical distribution in Boltzmann approximation [3]. This distribution was received by Boltzmann for ideal gases with the assumed statistical independence of particles. However, this assumption contradicts mechanics. The stochasticity generated by X-jumps substantiates it.

Let's show by the example of the system energy evaluation that probability of a state with energy $E_{(m)}$, received on the basis of one-partical distribution, gives correct value of the physical quantities characterizing the system as a whole. Let's represent the statistical sum (12) as follows:

$$Z = e^{-\beta A}, \qquad (13)$$

where A is the free energy of the system.

It is known that internal energy of a system is defined by:

$$E = \frac{\partial(\beta A)}{\partial\beta}$$

From equations (13) and (12) it is found that:

$$E = -\frac{\partial(\ln Z)}{\partial\beta} = -N\frac{\partial(\ln Z_1)}{\partial\beta} = N\frac{\sum_i \varepsilon_i e^{-\beta\varepsilon_i}}{Z_1} = \sum_i \varepsilon_i n_i, \qquad (14)$$

which corresponds to the initial formulation of the problem.

Now there is a question of canonical distribution taking into account the energy of interaction in the system. Hamiltonian of such a system can be written down in the following form:

$$\hat{H} = \hat{H}_0 + \hat{H}_{int}, \quad \hat{H}_{int} = \sum_{i,j} \hat{H}_{ij}, \qquad (15)$$

where $\hat{H}_{ij}$ is the interaction energy operator of the particles $i$ and $j$.

It is possible to consider various terms of interaction Hamiltonian as energy of some pseudo-particles. It is natural to assume that stochastisation in the system leads to statistical independence of these pseudo-particles. Then the probability of an interactions system state with energy $E_{int}^k$ looks like:

$$W_{int}^k = \frac{\exp(-\beta E_{int}^k)}{\sum_l \exp(-\beta E_{int}^l)}. \qquad (16)$$

The joint probability of finding a system without interactions in a state with the energy $E_0^i$, and a system with interactions in a state with the energy $E_{int}^k$, is:

$$W_{ik} = \frac{\exp[-\beta(E_0^i + E_{int}^k)]}{\sum_{i,k} \exp[-\beta(E_0^i + E_{int}^k)]}. \qquad (17)$$

Expression (17), obviously, represents canonical distribution for the system with interactions, received as before without the aid of an environment.

Quantum mechanics does not allow to consider separately a system without interactions and a system with interactions, as we have done deriving distribution (17). It is only owing to stochastisation, caused by the existence of X-jumps, or X-probability, that we can pass to the description of a system with interactions by means of distribution (17). The assumption of statistical independence of objects in the system is natural and, most likely, is a necessary condition to use the reduced description of a macrosystem by means of parameter $\beta$.

Experimental confirmation of stochastization of interactions in a macrosystem does exist. In [4], the establishment of spin temperature was considered by methods of nonequilibrium thermodynamics. Agreement with the experiment has been achieved thanks to the assumption of statistical independence of the pseudo-particles in the interactions system, and of the existence of a source of irreversibility beyond the equations of mechanics.

When the interaction between particles in a system or between its subsystems is weak, the equilibrium is established slowly. It is therefore natural to consider that the probability of X-jumps is not divorced from the energy characteristics of the system, and it grows with the growth of interaction energy in it. It explains both the hierarchy of relaxation times and the possibility to allocate subsystems with different temperatures within a system.

When the internal energy of the system and, accordingly, its temperature, decreases, there comes the moment when the subsystem of interactions can no longer be considered as a set of statistically independent pseudo-particles. New statistically independent objects are formed in the system, i.e. there is a phase transition. The study of this process from the point of view of the X-probability postulate could become an interesting and promising direction in statistical mechanics.

**Canonical distribution and quantum mechanics**

Remember that we did not need to resort to environment in order to derive the canonical distribution. We will show now that canonical distribution cannot be the consequence of a quantum mechanical evolution of a system. Indeed, a system's quantum states are found by the solution of the Schrödinger equation:

$$i\hbar\frac{\partial\psi}{\partial t} = \hat{H}\psi. \qquad (18)$$

A system state function can be written as a linear combination of eigenfunctions of the operator $\hat{H}$:

$$\psi = \sum_m a_m(t)\,\psi_m. \qquad (19)$$

The energy of the system in state (19) is:

$$E = \sum_m |a_m(t)|^2 E_m. \qquad (20)$$

For function (19) the Schrödinger equation is written as follows:

$$i\hbar\sum_m \psi_m \frac{\partial a_m(t)}{\partial t} = \sum_m \psi_m E_m a_m(t). \qquad (21)$$

Passing to the equations for coefficients $a_m$, we find:

$$i\hbar\frac{\partial a_m(t)}{\partial t} = E_m a_m(t), \qquad (22)$$

From equation (22) we have

$$a_m(t) = a_m(0)\, exp\left(-\frac{i}{\hbar}E_m t\right). \qquad (23)$$

$|a_m(t)|^2$ defines the probability of the system to be in a quantum state with energy $E_m$. As $|a_m(t)|^2 = |a_m(0)|^2$, we see that the quantum mechanics does not allow the system to pass into a state with a set of $|a_m(t)|^2$ different from the initial. Hence, from the point of view of quantum mechanics, the use of the canonical distribution (2) is impossible. It once again confirms that processes must exist, which provide for stochastisation and allow to use probabilistic methods for the description of macrosystems behavior.

**Mixing and separation of temporal correlation**

The X-jumps also determine the mixing character of macrosystems evolution.
Let us denote

$$Z(t) = (q(t), p(t)), \tag{24}$$

as a point in the phase space which characterizes the system state. Let us also denote

$$\langle f \rangle = \int f(Z)\rho(Z)dZ \tag{25}$$

as a phase space average of an arbitrary integrable function of $Z$.

Let us define the correlation function $R(f, g|T)$. If $f$ and $g$ are two arbitrary functions of $Z(t)$, then

$$R(f, g|T) = \langle f[Z(t+T)]g[Z(t)]\rangle - \langle f[Z(t)]\rangle\langle g[Z(t)]\rangle \tag{26}$$

The phase density $\rho(Z)$ is independent of $t$. Therefore, the phase space averages in equation (26) are also independent of $t$.

The value of the function $g(Z)$ within the boundaries $\rho(Z)dZ$ may be considered constant. After the time $T$ passes, the values of the function $f(t + T$ will be determined by the position of the points $Z(t+T)$ of the phase space, departed from the element $\rho(Z)dZ$. If for the time $T$ every phase drop $\rho(Z)dZ$ gets spread all over the phase space and fills it with the density proportional to the density $\rho(Z)$, then every value $g(Z)$ in equation (26) will have a factor which is the average $\langle f \rangle$ with the weight proportional to the phase density $\rho(Z)$. Hence the integral over the whole phase space in equation (26) gives the product of the average $\langle f \rangle$ by the average $\langle g \rangle$. In this case the system evolution is called mixing. In physics the satisfaction of the condition

$$\lim_{T\to\infty} R(f, g|T) = 0 \tag{27}$$

is called separation of temporal correlations.

Obviously for the spreading of the phase drop $\rho(Z)dZ$ over the whole phase space it is necessary that the nearby points in $\rho(Z)dZ$ diverge in the course of time far away from each other. In other words, non-stability of the system evolution is required. But for description of real systems, which consist of microparticles, one should use the quantum mechanics equations (in this case the role of the phase space is played by the system states space or the space of coefficients in the expansion of the state function in terms of the eigenfunctions). These equations have no exponentially diverging solutions and, therefore, the corresponding systems are void of local non-stability necessary to satisfy the mixing condition. The connection between the mixing in quantum systems and inaccuracy of quantum mechanics was discussed in the paper [1], devoted to time reversal in spin systems. However, the character of this inaccuracy was not specified in that paper.

Only X-jumps can make the phase drop to spread over the whole space of the system states, which results in separation of temporal correlations. Note that, when the mixing condition is satisfied, it follows that the ergodicity condition is automatically satisfied too.

**Conclusion**

So, X-jumps determine the following: (1) stochastization of a system and occurrence of irreversibility in it; (2) meeting the condition of mixing and ergodisity; (3) equality of observed and ensemble average values of physical quantities. In other words, the use of the X-jumps postulate allows, from our point of view, to solve at last the problem of substantiation of statistical mechanics, and at the same time to understand why tomorrow differs from yesterday. Besides, relieving us from mechanical determinism, the X-probability promotes freedom of our will.

In connection with aforesaid, the study of X-jumps becomes significant. With reduction of observation time and amount of particles in the system, the measured values will deviate from the averages on ensemble. Estimation of the X-probabilities could be accessed through experiments on systems with small number of particles where solution of the Schrödinger equation is possible. Computer experiments on model systems, introducing randomness in addition to the mechanical evolution of system, could also be interesting (at present, such experiments are achievable for a sufficiently large number of particles). By means of these experiments it is possible to study a correlation between the number of particles, the probability of X-jumps and the observation time sufficient for a reduced description of the system by means of canonical distribution.

We feel that the introduction of the X-probability postulate is a necessary and long overdue statement of the fact (note that all working principles and equations in physics are nothing more than a successful statement of the fact). We will be quite happy if existence of the X-probability begins to be taken for granted. In any case, we cannot imagine a simpler assumption which could make it possible to put things in their places without contradicting the existing picture of the world. Nature does not like to make things difficult. Our matter is to make the right guess.

**References**

1. Skrebnev, V.A. & Safin, V.A. Experimental study of reversibility and irreversibility in the evolution of the nuclei spin system of $^{19}F$ in $CaF_2$. *J. Phys. C: Solid State Phys.* **19**, 4105–4114 (1986).
2. Skrebnev, V.A. & Zaripov, R.N. Investigation of the equilibrium establishment in the spin system by using dipole magic echo. *Appl. Magn. Reson.* **16**, 1- 17 (1999).
3. Balescu, R. *Equilibrium and Nonequilibrium Statistical Mechanics* (AWiley-
Interscience Publication, New York, 1975).
4. Skrebnev,V.A. Dipole magic echo in thermodinamical systems. *J. Phys.: Condens. Matter* **2**, 2037–2044 (1990)